# Low Temperature Performance of a Large Area Avalanche Photodiode

V.N. Solovov* , F. Neves, V. Chepel, M.I. Lopes, R. F. Marques and
A.J.P.L. Policarpo

LIP-Coimbra and Department of Physics of the University of Coimbra,
3004-516 Coimbra, Portugal

**Abstract**

A Large Area Avalanche Photodiode (LAAPD) was studied, aiming to access its performance as light detector at low temperatures, down to –80°C.

The excess noise factor, F, was measured and found to be approximately independent of the temperature. A linear dependence of F on the APD gain with a slope of 0.00239±0.00008 was observed for gains >100.

The detection of low intensity light pulses, producing only a few primary electron-hole pairs in the photodiode, is reported.

PACS: 85.60.Gz, 85.60.Dw, 70.70.+m

Keywords: photodetectors, excess noise factor, avalanche photodiode, LAAPD,

* Correponding author: solovov@lipc.fis.uc.pt

## 1. Introduction

Large Area Avalanche Photodiodes (LAAPD) are of great interest for high efficiency detection of low intensity light pulses from scintillators. In spite of some drawbacks, they can compete successfully with photomultiplier tubes (PMT) in many applications that require very compact detector packaging (in Positron Emission Tomography, for instance), low sensitivity to magnetic fields (as in some space applications and accelerator experiments) or low intrinsic radioactivity as required for the low background experiments (e.g., search of dark matter). Very good time and energy resolutions have been achieved with LAAPDs with various scintillators, namely inorganic crystals [1, 2] and liquid xenon [3].

Operation at low temperature can be not only a necessity, as in the case of liquid xenon [3], but also a favorable option. In fact, to operate the LAAPD at low temperature has two advantageous effects: 1) the dark current noise decreases dramatically; 2) the voltage required to achieve a certain gain becomes lower. In our previous work [4], both effects were investigated as a function of the temperature between 25 °C and -100 °C.

In a detector with a LAAPD readout, the noise and the fluctuations of the avalanche gain contribute to the energy resolution of the system. The later is usually expressed in terms of the excess noise factor, F, [5]. Previous measurements of the energy resolution in liquid xenon excited with alpha-particles [3], indicated a value of F significantly larger than the value usually reported for room temperature.

In this paper, we report on measurements of the excess noise factor as a function of the gain at low temperatures (down to –80°C). The observation of very low intensity light pulses that produce only a few primary electron-hole pairs in the photodiode is also presented.

## 2. Experimental Set-Up

In our measurements we used a windowless LAAPD, 5 mm in diameter, from Advanced Photonix, Inc [6]. This photodiode, manufactured using beveled-edge technology, has low dark current and relatively small capacitance (typically, about 30 nA and 25 pF, respectively, at room temperature and a gain of 200).

The set-up used for the measurements is schematically depicted in Fig. 1. The APD connected to a low-noise charge sensitive preamplifier (Cremat CR-101D) was mounted inside a metallic cage and placed into a liquid nitrogen cryostat.

Light pulses of 0.5 μs in duration were produced by a green LED driven by a pulser. The LED was mounted in a black box outside the cryostat and optically connected to the APD through an optic fiber. The intensity of light pulses generated by the LED was monitored by a PMT (Hamamatsu R1668). The output signal of the APD was amplified by a charge sensitive preamplifier followed by a spectroscopy amplifier (Canberra 2021). Semigaussian shaping with shaping time of 1 μs was used. The amplified and shaped signals from both the APD and the PMT were digitized by a peak ADC (LeCroy 2259B) and stored in a PC-based multichannel analyzer. The PC also controlled the LED and calibration pulsers and, synchronously, provided a gate for the ADC. The linearity, amplification, pedestal (offset) and the electronic noise of the whole spectrometric channel were measured using calibration pulses from a high precision pulser fed to the input of the preamplifier through a capacitance of 2.2 pF. The calibration was verified by observing the pulse height spectrum due to the direct conversion of 60 keV γ-rays, from an $^{242}$Am source, in the photodiode at unitary gain.

For precise measurement of the excess noise factor it is very important to maintain the stability of the APD gain during the acquisition. This requires stable bias voltage and temperature as the gain strongly depends on both of them, especially at high gain values.

In our measurements, the stability of the bias voltage was better than 0.1 V. To reduce the ripples from the HV supply (CAEN N471), an RC-filter with a time constant of 1 s was used. The variations of the temperature were measured with a precision of 0.02°C using a platinum thermoresistor fixed directly to the photodiode. During the acquisition of a single data point, the temperature variation did not exceed 0.2°C.

## 3. Experimental Methods and Results

### *A. Measurement of Gain*

The gain of the photodiode as function of the bias voltage was measured operating the LED in pulse mode. For each value of the bias voltage, the distribution of the amplitude of the APD charge signals was recorded. The mean value of the distribution was normalized to the mean amplitude of the signals measured at low voltage ($V_{bias}$=400V) where the gain is unitary [7]. Some results are plotted in Fig. 2.

### *B. Measurement of Excess Noise Factor*

In an avalanche photodiode, the random character of the multiplication process of the electron-hole pairs under an electric field introduces the so-called multiplication noise, which is characterized by the excess noise factor, *F*, defined as

$$F = \frac{\langle m^2 \rangle}{\langle m \rangle^2} = \frac{\langle m^2 \rangle}{M^2} \qquad (1)$$

where *m* is the gain of an avalanche in the photodiode and $M = \langle m \rangle$ [8].

By considering the statistical nature of the avalanche formation, it was shown that F can be approximated to

$$F \approx k_{eff}M + \left(2 - \frac{1}{M}\right)(1 - k_{eff}) \qquad \text{for M>>1} \qquad (2)$$

where $k_{eff}$ is the effective ionization rate ratio [9].

The fluctuations of the amplitude of the charge signal of an APD illuminated with light pulses can be expressed as [5]:

$$\frac{\sigma^2}{A^2} = \left(\frac{\sigma_n}{MN_o}\right)^2 + \frac{F-1}{N_o} + \delta^2 \qquad (3)$$

where $A \equiv MN_0$ is the average amplitude of the pulses at the photodiode output expressed in number of electrons, $N_0$ the average number of primary electron-hole pairs created by a single light pulse, $\sigma^2$ the variance of the signal amplitude distribution (in number of electrons), $\sigma_n$ the electronic noise at the input of the preamplifier (r.m.s. in number of electrons), and $\delta^2$ is the relative variance in $N_0$ and can be written as

$$\delta^2 = \frac{1}{N_0} + \varepsilon^2 \qquad (4)$$

where $\varepsilon^2$ includes all the non-Poisson contributions to the fluctuations in $N_0$. In the present measurements, the fluctuations of the intensity of the light source and the gain non-uniformity in the photodiode can contribute to $\varepsilon^2$. Replacing eq. (4) in (3), one gets

$$\sigma^2 = \sigma_n^2 + M^2 N_0 F + \varepsilon^2 M^2 N_0^2, \qquad (5)$$

If the light source is stable and the gain non-uniformity are not relevant such that $\varepsilon << \sqrt{F/N_0}$, the last term in (5) can be neglected (in our measurements the worst case corresponds to $F \approx 2$ and $N_0 \approx 5000$, which gives us condition $\varepsilon << 0.02$) and the excess noise factor determined from the expression:

$$F = \frac{\sigma^2 - \sigma_n^2}{M^2 N_0} \tag{6}$$

or, more conveniently,

$$F = \frac{\sigma^2 - \sigma_n^2}{M^2 N_0^2} N_0 = \frac{\sigma^2 - \sigma_n^2}{A^2} N_0 \tag{7}$$

For measuring the excess noise factor, the amplitude distribution of the APD output, as well as the distribution of the calibration pulses, were acquired for different values of the gain under fixed intensity of the light pulses. A typical pulse height spectrum is shown in Fig. 3. From these distributions, the following values were determined: $\sigma^2$ and A as the variance and the mean of the former distribution and $\sigma_n^2$ as the variance of the distribution of the calibration pulses. The value of $N_0$ was measured at M=1 as the mean charge signal amplitude expressed in number of electrons. It was set between 500 and 4000 depending on the run, and was kept fixed during each run.

These measurements were carried out for gains up to 2000 in the temperature range from 0°C down to –80 °C. The results for –40 °C and –80 °C for M>100 are represented in Fig. 4. In all the cases, the excess noise factor increases linearly with gain, i.e.,

$$F = F_0 + kM \tag{8}$$

No temperature dependence of $F_0$ and k was observed within the experimental errors (Fig.5). Averaging the values of $F_0$ and k attained at different temperatures, one gets $F_0 = 1.87 \pm 0.02$ and $k = 0.00239 \pm 0.00008$. These results are consistent with those presented in [10].

Fig. 6 shows the noise of the photodiode and preamplifier, $\sigma_n$, as a function of the APD gain for different temperatures. For $T \leq -40$°C and $M \geq 5$, the noise is approximately constant and ≈ 250 electrons (r.m.s.). At higher temperatures (T=-20°C), the noise increases dramatically for large gains due to the increase of the dark current.

### *C. Low-intensity signal detection*

The reduction of the dark current at low temperature allows the detection of light signals of very low intensity, which produce just a few primary electron-hole pairs in the photodiode. In Fig.7, it is shown the spectrum corresponding to $N_0$=4.3, $N_0$ being the mean number of electron-hole pairs per light pulse (we observed similar spectra for $N_0$ varying from 3 to 7 electron-hole pairs). For these measurements, the calibration in $N_0$ was carried out with the photodiode gain equal to 1 and using the PMT signal as reference (see Fig.1).

To avoid possible non-linearity in the first channels of the ADC, an offset was added by sending a test pulse of fixed amplitude and duration at the preamplifier input simultaneously with the light pulse. The value of the offset was equivalent to 1500 electrons at the photodiode output. Hence, the broad pulse distribution in Fig.7 corresponds to the offset light pulses while the sharp peak at the left is due to the test pulses and is acquired with the LED turned off. The upper axis in Fig.7 was scaled to the number of primary electron-hole pairs by dividing the charge measured at the photodiode output by the APD gain, M. Similar spectra were observed in [10].

## 4. Conclusions.

The excess noise factor of a Large Area Avalanche Photodiode was measured as a function of gain and temperature. For the APD gains >100, it was found to be linear with gain and practically independent on temperature in the range from 0°C down to -80°C. The amplitude distributions of the signals with the average number of primary electron-hole pairs as small as 3 to 7 were measured, thus showing the possibility of detection of very low intensity light pulses.

**Acknowledgements**

This work was financed by the project CERN/FNU/43729/2001 from the Fundação para a Ciência e Tecnologia, Portugal.

**Figure Captions**

Figure1: The experimental set-up for studying low-temperature performance of APDs: PA – charge sensitive preamplifier, A – spectroscopy amplifier, GG – gate generator, F – RC filter.

Figure 2: Dependence of the APD gain on the bias voltage at different temperatures.

Figure 3: A typical pulse height spectrum due to the LED (at the right) and calibration pulses (at the left). The horizontal axis shows the charge at the preamplifier input. Temperature is -60°C, APD gain is 33, $N_0$ is equal to 4000 electrons.

Figure 4: Dependence of the excess noise factor on the APD gain at -40°C and -80°C. The straight line $F=F_0+kM$ is fitted to the experimental results with $F_0$ and k as adjusted parameters.

Figure 5. $F_0$ and k obtained at different temperatures.

Figure 6. Dependence of the electronic noise at the preamplifier input sensitive on the APD gain and temperature (the preamplifier was connected to the APD).

Figure 7. A spectrum of the APD output signals corresponding to a mean number of primary electron-hole pairs equal to 4.3. Temperature is –40°C, APD gain is 600.

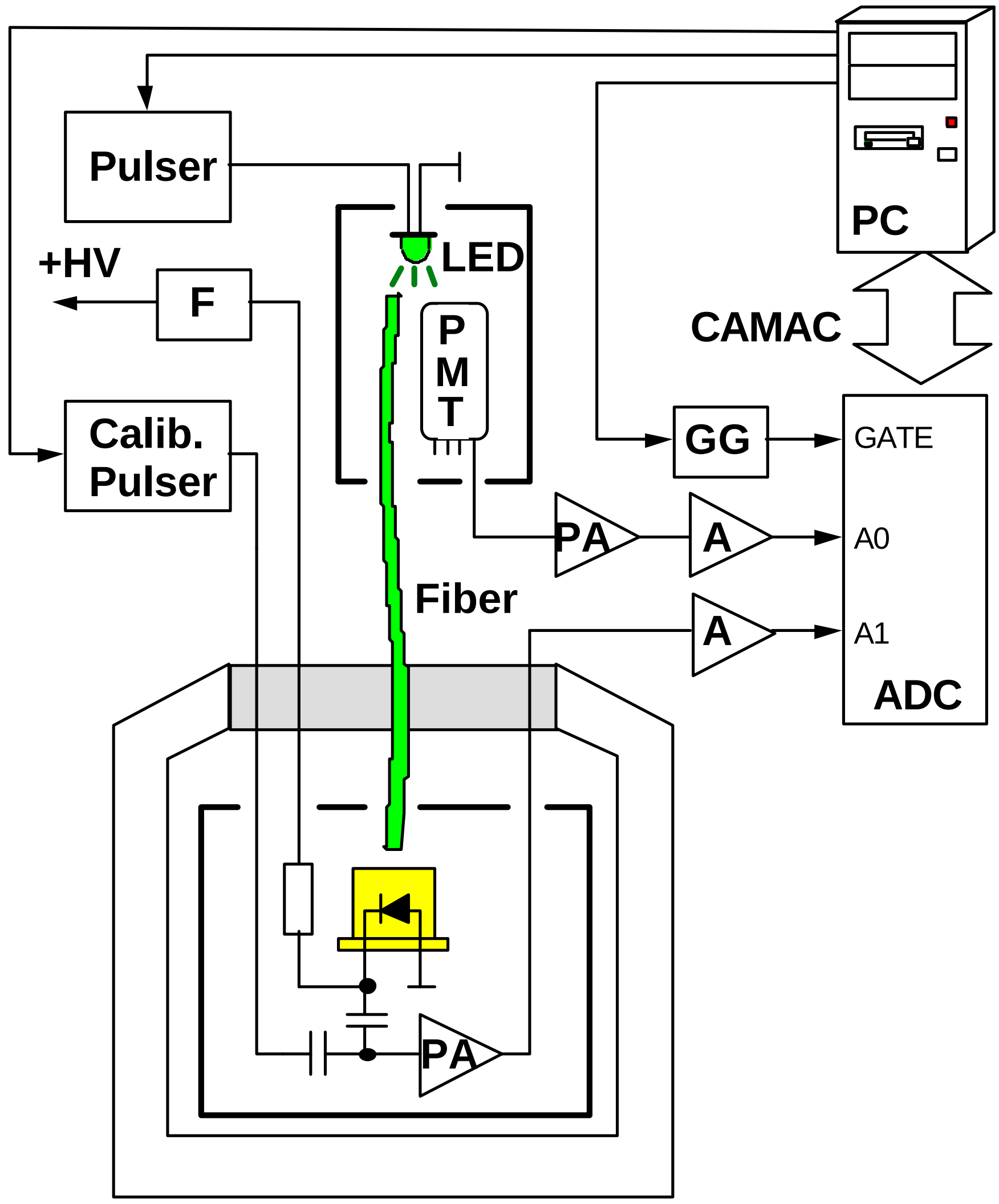

Figure 1

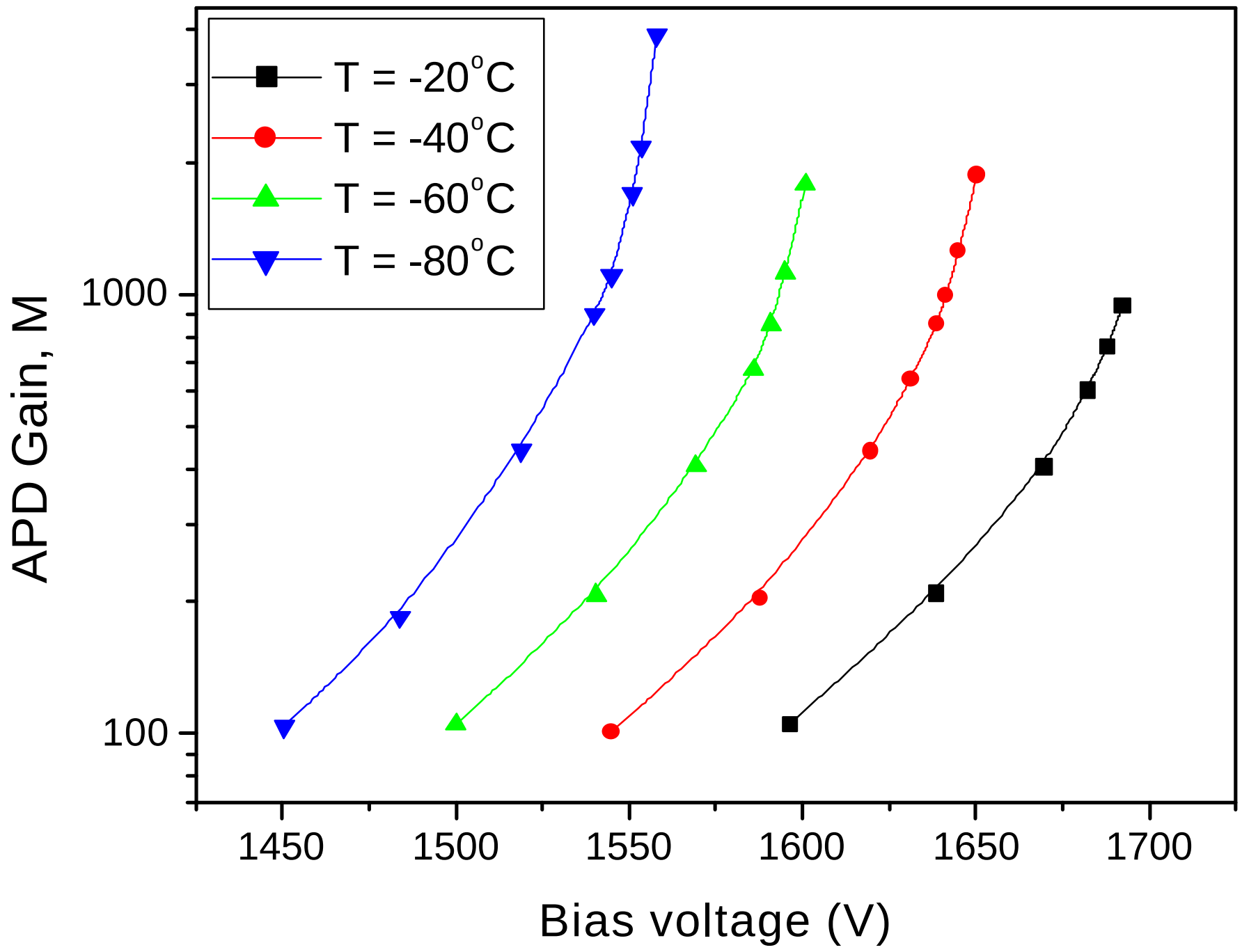

Figure 2

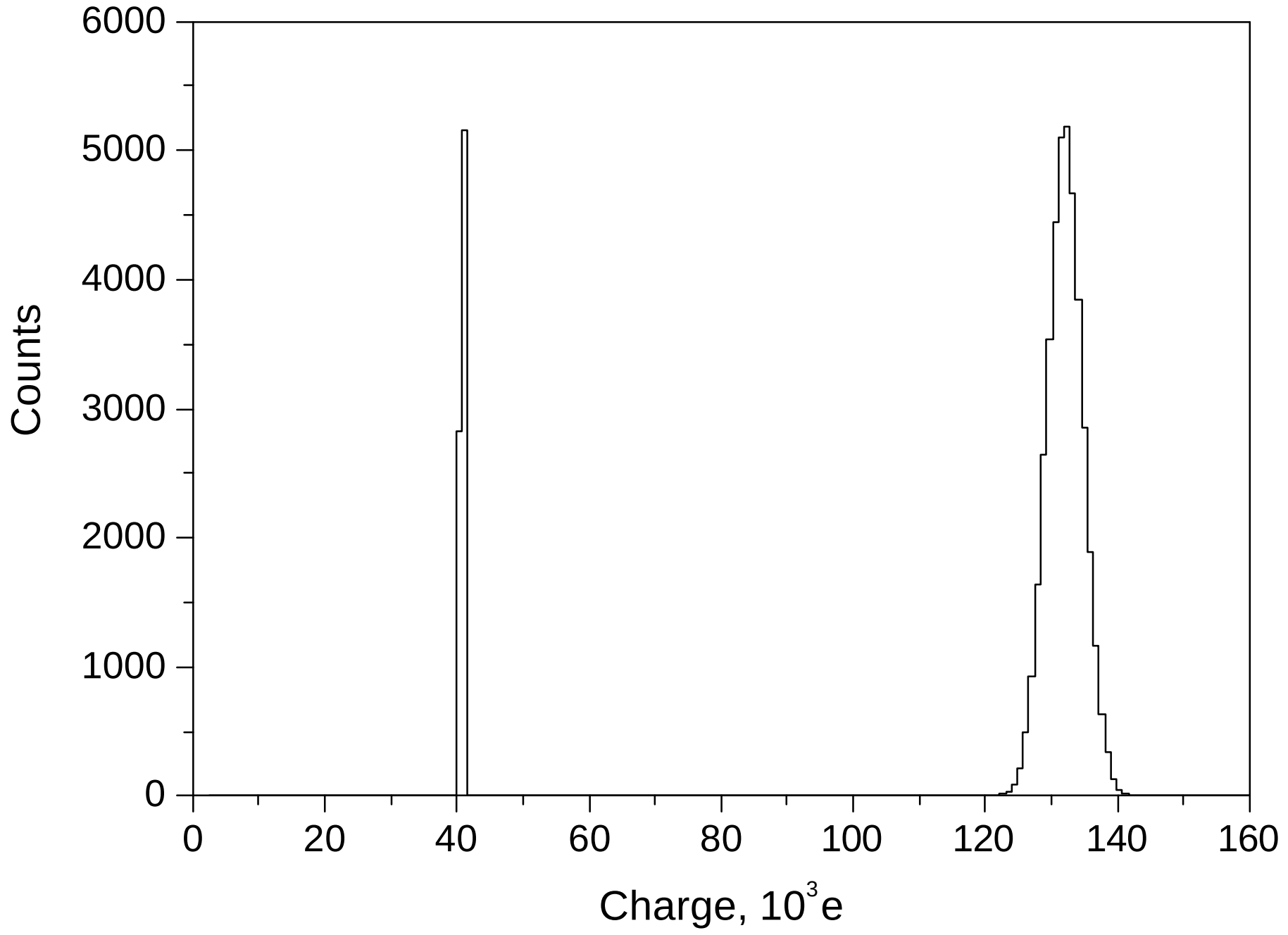

Figure 3

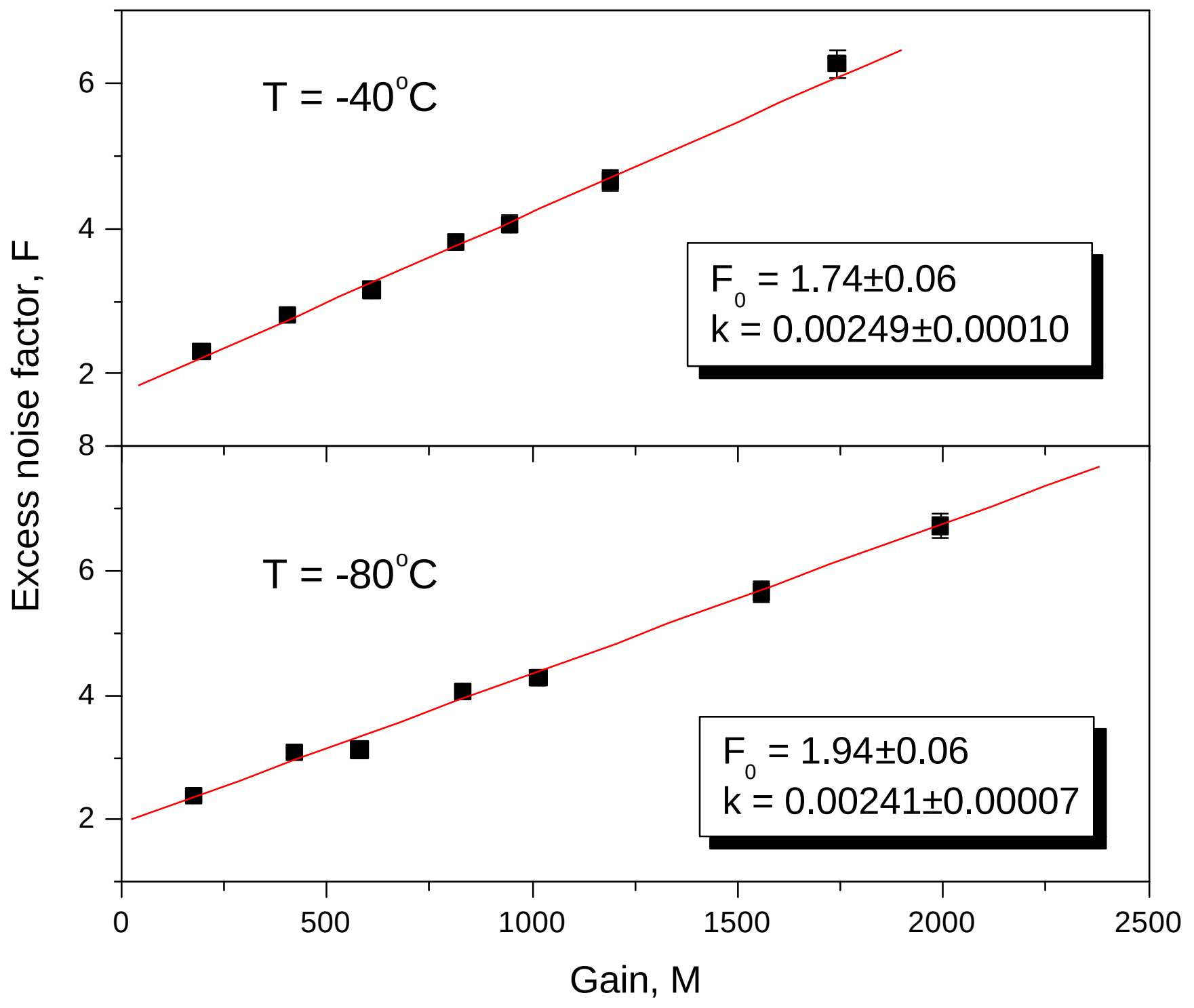

Figure 4

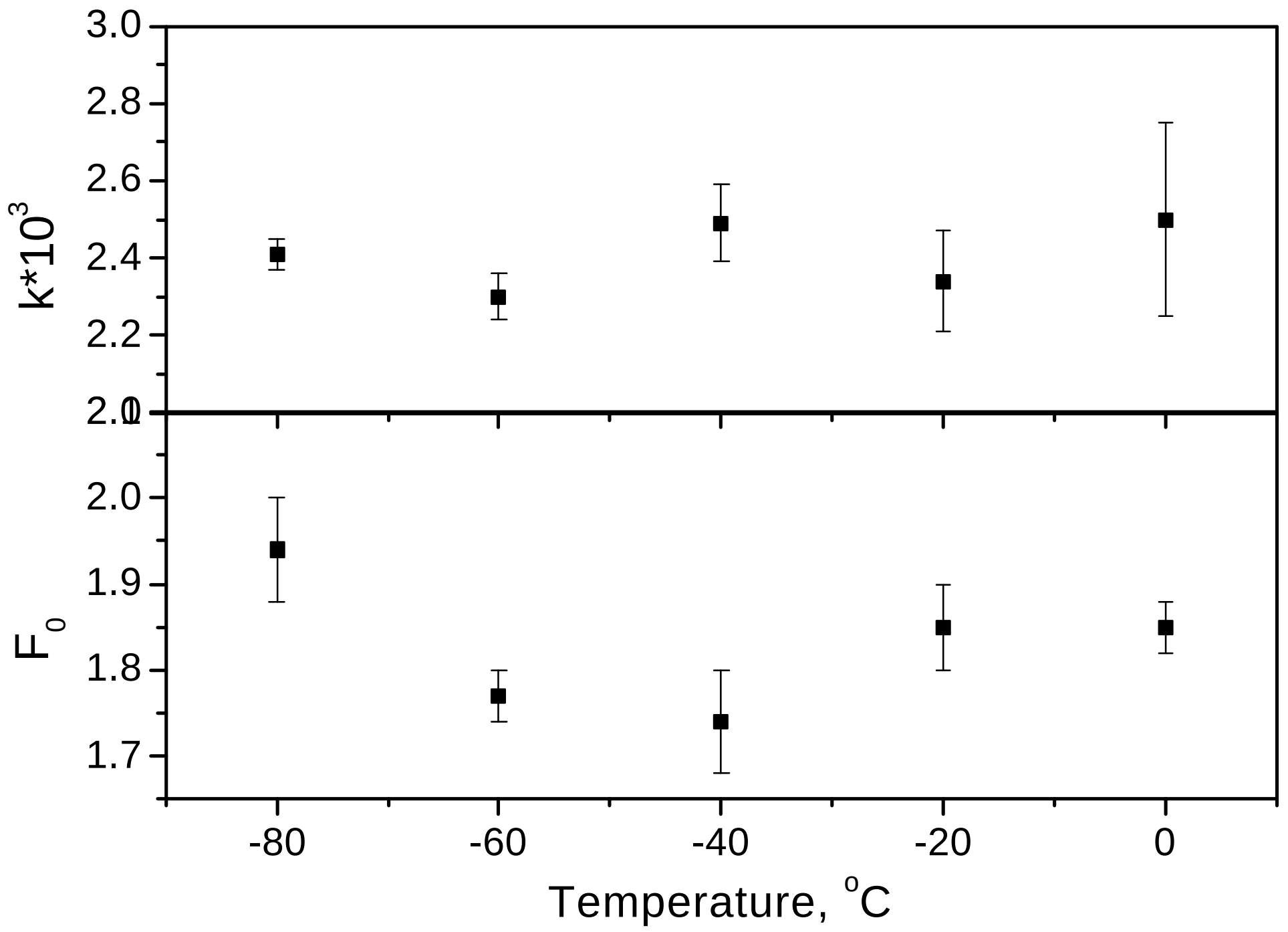

Figure 5

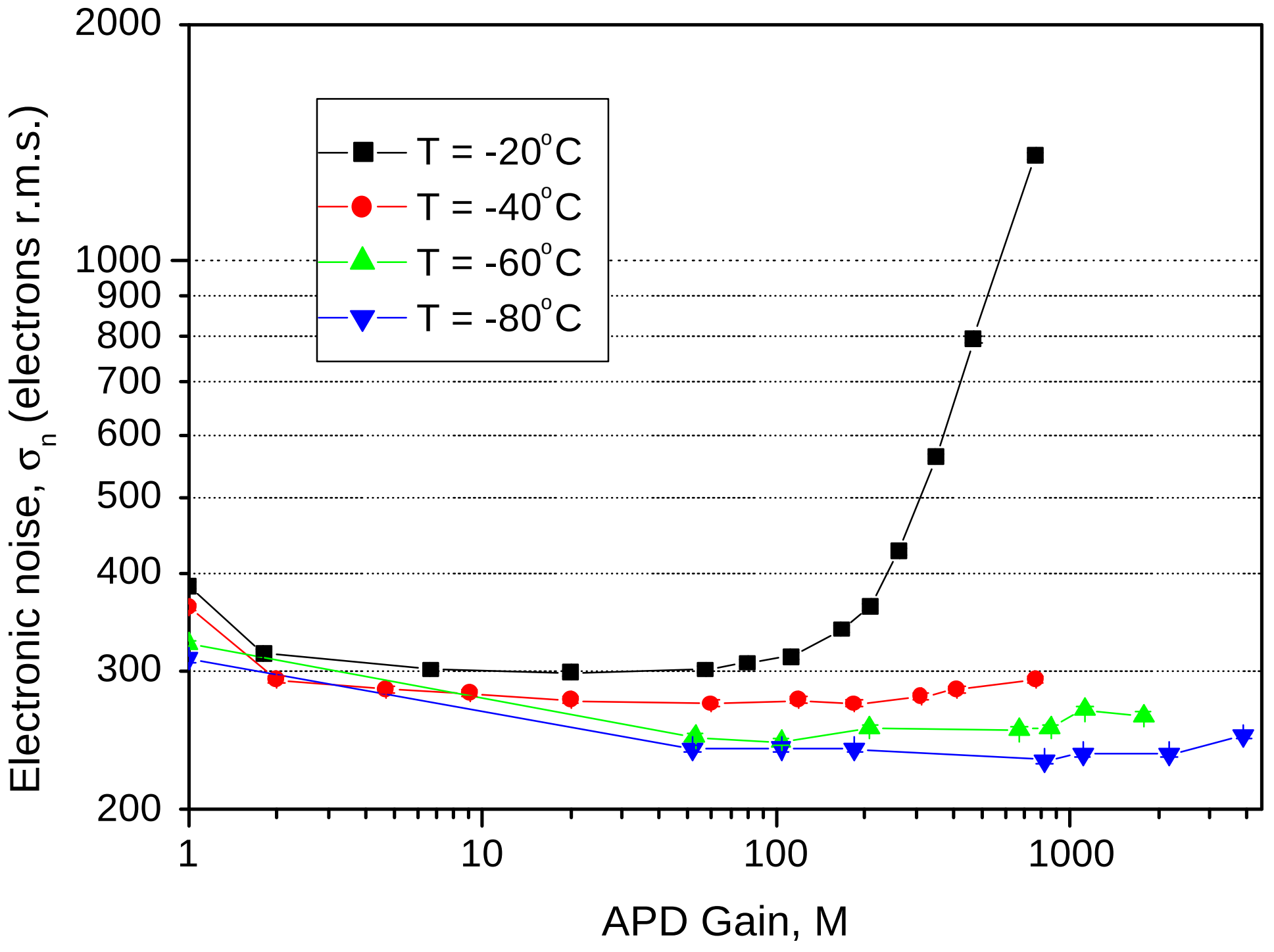

Figure 6

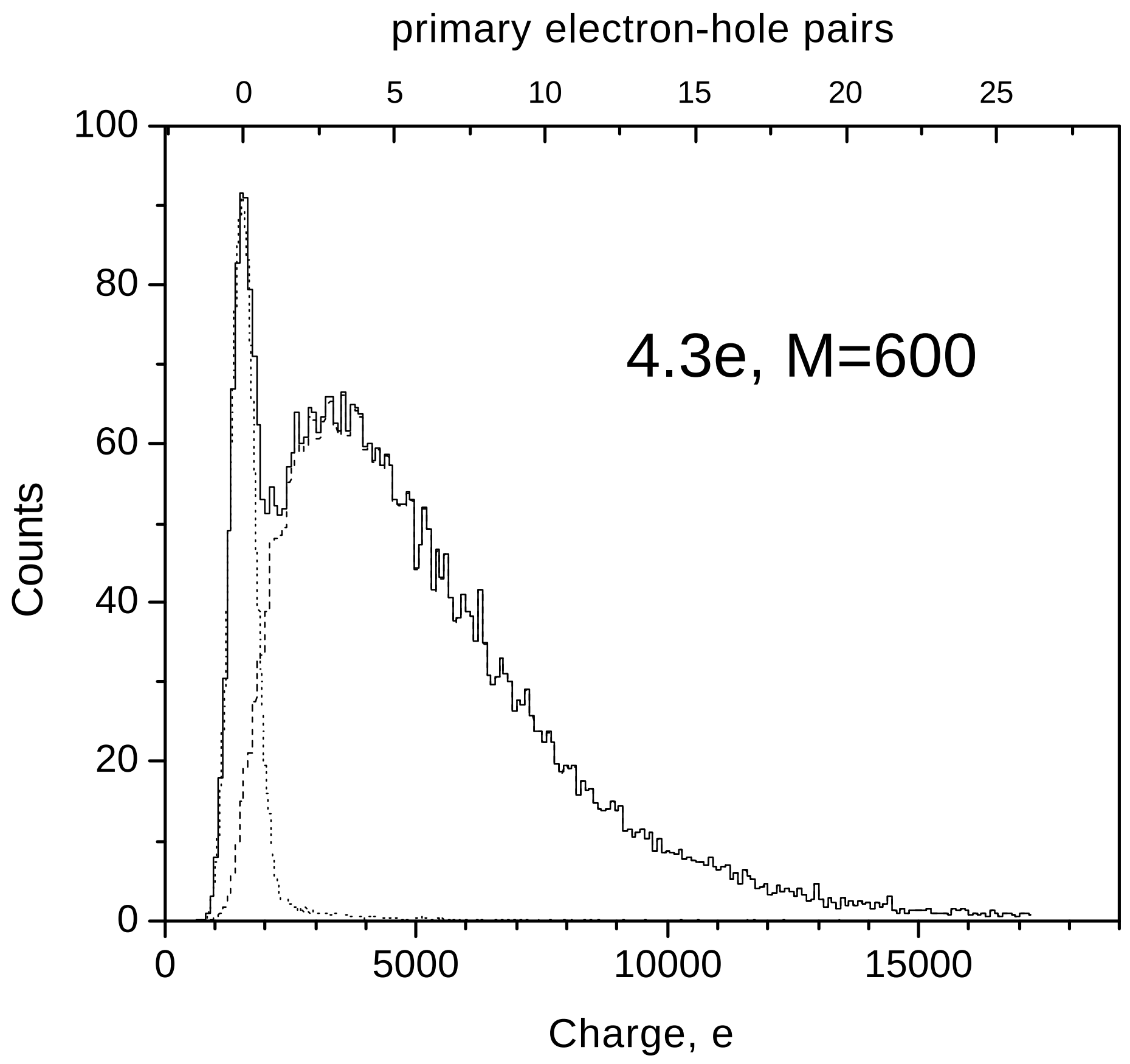

Figure 7